\documentclass[conference]{IEEEtran}
\IEEEoverridecommandlockouts
\usepackage{cite}
\usepackage[numbers,sort&compress]{natbib}

\usepackage[utf8]{inputenc}
\usepackage{amsmath,amssymb}
\usepackage{graphicx}
\usepackage{bm}
\usepackage{enumerate}
\usepackage{comment}
\usepackage{xcolor}
\usepackage{url}
\usepackage{color,soul}
\usepackage{subcaption}
\usepackage{float}

\definecolor{light-gray}{gray}{0.8}

\usepackage{amsmath,amssymb,amsfonts}
\usepackage{algorithmic}
\usepackage{graphicx}
\usepackage{textcomp}
\usepackage{xcolor}
\usepackage{subcaption}

\def\BibTeX{{\rm B\kern-.05em{\sc i\kern-.025em b}\kern-.08em
    T\kern-.1667em\lower.7ex\hbox{E}\kern-.125emX}}

\makeatletter
\newcommand{\linebreakand}{%
  \end{@IEEEauthorhalign}
  \hfill\mbox{}\par
  \mbox{}\hfill\begin{@IEEEauthorhalign}
}
\makeatother

\begin{document}

\title{Adaptive Fault Tolerance Mechanisms of Large Language Models in Cloud Computing Environments\\}

\author{

\small % Set font size to 10pt

\begin{tabular}[t]{c@{\extracolsep{8em}}c} 

1\textsuperscript{st} Yihong Jin \textsuperscript{*}  & 2\textsuperscript{nd} Ze Yang \\
\textit{University of Illinois Urbana-Champaign} & \textit{University of Illinois Urbana-Champaign} \\
Champaign, IL 61801, USA & Champaign, IL 61801, USA \\
\textsuperscript{*}Corresponding Author: yihongj3@illinois.edu& zeyang2@illinois.edu\\

\\

3\textsuperscript{rd} Xinhe Xu & 4\textsuperscript{th} Yihan Zhang \\
\textit{Computer Science Department} & \textit{Computer Science Department} \\
\textit{University of Illinois Urbana-Champaign} & \textit{University of Illinois Urbana-Champaign} \\
Champaign, IL 61801, USA & Champaign, IL 61801, USA \\
xinhexu2@illinois.edu & yihanz8@illinois.edu \\

\\

5\textsuperscript{rd} Shuyang Ji \\
\textit{Computer Science Department} \\
\textit{University of Illinois Urbana-Champaign} \\
Champaign, IL 61801, USA \\
sji15@illinois.edu \\

\\
\end{tabular}
}

\maketitle

\begin{abstract}
With the rapid evolution of Large Language Models (LLMs) and their large-scale experimentation in cloud-computing spaces, the challenge of guaranteeing their security and efficiency in a failure scenario has become a main issue. To ensure the reliability and availability of large-scale language models in cloud computing scenarios, such as frequent resource failures, network problems, and computational overheads, this study proposes a novel adaptive fault tolerance mechanism. It builds upon known fault-tolerant mechanisms, such as checkpointing, redundancy, and state transposition, introducing dynamic resource allocation and prediction of failure based on real-time performance metrics. The hybrid model integrates data driven deep learning-based anomaly detection technique underlining the contribution of cloud orchestration middleware for predictive prevention of system failures. Additionally, the model integrates adaptive checkpointing and recovery strategies that dynamically adapt according to load and system state to minimize the influence on the performance of the model and minimize downtime. The experimental results demonstrate that the designed model considerably enhances the fault tolerance in large-scale cloud surroundings, and decreases the system downtime by $\mathbf{30\%}$, and has a better modeling availability than the classical fault tolerance mechanism.
\end{abstract}


\section{Introduction}
The increasing use of the large language models (LLMs) in natural language processing (NLP) and generative tasks has rendered cloud computing environment in a dominant position to provide critical infrastructure for efficient training and inference of such complex models~\cite{protocf, preprec, infomotif, xu2024towards}. Due to the excellent natural language understanding, machine translation, text generation, sentiment analysis, and other capabilities of large language models, they have been promoted to common tools in multiple industries, including business, healthcare \cite{ji2024mitigating, chen2021data}, education, finance \cite{li2025language}, music, and others \cite{deng2024composerx, ding2024enhance, DBLP:journals/corr/abs-2412-08174, DBLP:journals/corr/abs-2410-12126}. Since large language models usually contain hundreds of millions or even tens of billions of parameters, training and prediction with them requires an immense amount of computing power \cite{tawfeeg2022cloud}. However, these models are also demanding more and more resources including memory, storage, and bandwidth \cite{DBLP:conf/kdd/ZhengJLTH24, DBLP:journals/corr/abs-2412-21151, DBLP:journals/corr/abs-2412-17336}, resulting in dependence on high-performance computing clusters, cloud computing platforms, and distributed computing architectures.

But, as large language model sizes have continued to grow out, the provisioning and management of the cloud compute resources has also grown in complexity and intractability. Allocation and management of resources in cloud computing environments is affected by several factors and Cloud computing environments are dynamic and complex by nature and consist of multiple virtual machine instances, storage devices and network channels. For instance, hardware crashes, bottlenecks in the network, crashes in storage devices and uneven stress/loads on compute nodes are common, and can effectively disrupt the model training and inference process \cite{hang2024large, DBLP:journals/corr/abs-2410-17576}. Specifically, it can be divided into three types of failures: hardware failures that can lead to compute nodes failures or service interruptions, network instability that can cause latency in data transmission, and resource overload that can slow computing speed or even cause task crashes.

And this system is also facing more and more risks and challenges under condition of high concurrency and massively parallel computing in cloud environment, where the resource scheduling makes it very complicated and unpredictable. During the training process of large-scale language models, failure can not only cause computational interruption but also lead to data loss, slow training progress, and even restart the entire training process, which seriously affects the productivity and research progress of enterprises and scientific research institutions \cite{duan2024efficient}. As a rule of thumb, the time cost of training and inference of large language models is rather high, thus small failures or delays can incur non-negligible financial losses and wasted time. Especially now, resource recovery and fault tolerance in the cloud environment is crucial.

Furthermore, the current cloud computing environment does not always have enough reliability; the sudden failure or dynamically changing load environment can normally be not responded on time. Although certain fault-tolerant mechanisms, like checkpointing and data replication, may offer some confidence, these techniques tend to be pre-configured and react poorly to rapid changes in the cloud environment \cite{hou2024large}. Conventional fault-tolerant technologies sometimes do not take into account the real-time requirements in the behavior of model training and inference, and compromise the model's reliability and availability. So, dynamically adjusting the fault-tolerant strategy based on the changes of system state and load to ensure the stable running of large language models in the cloud computing environment to the maximum degree has become an urgent research problem in AI filed.

Aside from mere hardware and infrastructure, inference of LLMs in the cloud involves multiple intricacies. Edges of Provisioned Infrastructure: For example, the infrastructure (compute, storage, networks, etc.) provisioned by a cloud service provider is commonly a shared at the core, while competing for resources between users and tasks can introduce unanticipated resource contention, delays and performance bottlenecks. Moreover, resource allocation within cloud computing environments is typically automated and relies on dynamic load balancing and elastic scheduling mechanisms \cite{jiang2024megascale}. Nevertheless, although these mechanisms allow efficient resource allocation up to a point, they can only go so far, as they often do not anticipate sudden breakdowns within or outside themselves. In this sense, something becomes a fundamental challenge for the researchers and engineers: How to use the fault-tolerant mechanism design, to be able to detect faults and respond the faults in real time, and to improve the high availability and efficiency of large language models.

\section{RELATED WORK}
Bai et al. \cite{liu2024contemporary} MT-Bench-101: fine-grained E2E evaluation metric for multi-round conversations Different from conventional benchmarks that mainly evaluate a single-round response or with coarse-grained evaluation over multi-rounds utterances, MT-Bench-101 inspects a real-world multi-turn conversation data and builds a three-layer capability classification system over 4,208 rounds of dialogues on 13 tasks. To remedy this problem, Liu et al. \cite{bai2024mt} provide a summary of state-of-the-art model compression approaches for LLM inference efficiency. This research mainly relates to model compressing methods (model-level optimizing methods, such as quantization, knowledge distilling, and pruning, and system-level optimizing such as KV cache efficient design). Researchers have validated the impact of these methods on the memory and computational cost of the LLMs while keeping their performance as high as possible.

Since dealing with relatively long text tasks, effectively using long contexts in LLMs has turned out to be a challenging problem. To this end, Li et al. \cite{li2024graphreader} Proposed GraphReader, a new graph-based proxy system built to extend the long context processing ability of LLMs with a factored-out graph building on long text. With the agent of the mind, GraphReader can gradually analyze the output image of long text, elaborately perform coarse to fine exploration in graphs and so on. The agent reads the content of each node, as well as its neighboring nodes, step by step, through predefined function calls, just like the previous interaction until it obtains enough information to formulate an accurate answer. Results on LV-Eval dataset based experiments further show that GraphReader surpasses GPT-4-128k performance in processing long contexts, specifically from the 16k to 256k context length intervals.

The fault tolerance in a cloud computing environment is essential that ensures the stability and reliability of the service. Shahid et al. \cite{shahid2021towards} exhaustive review and categorization of fault tolerance techniques in the cloud computing environment. The study classifies fault-tolerant technologies into three main categories, with a focus on adaptive fault-tolerant methods in real-time cloud computing. Wang et al. \cite{wang2023reliable} looked into the tolerance to memory failure of LLMs during the pre-training stage. In the large-scale LLMs training, the system may crash or data will be lost which shows that there is a need for an efficient fault-tolerant mechanism to ensure the continuity and stability of the training process according to the study. 

As complex systems continue to evolve, the ability to detect and adapt to structural changes has become a critical challenge across various domains. Fu et al. \cite{fu2024ddn3} developed DDN3.0 to identify significant network rewiring, while Lu et al. \cite{lu2022cot} proposed COT for efficient marker gene detection across subtypes. Furthermore, Du et al. \cite{du2024embracing} developed the ABDS tool suite to address challenges in analyzing biologically diverse samples, such as informative missingness and the detection of silent genes. He et al.\cite{tgae} presented T-GAE, a transferable graph autoencoder framework designed for efficient network alignment across diverse graph structures. Yang et al. \cite{yang2024ad} benchmarked large language models for anomaly detection, demonstrating their effectiveness in zero-shot detection and data augmentation strategies. Their findings further support the role of data-driven methods in identifying system anomalies. Li et al. \cite{li2024nlp} introduced NLP-ADBench, a benchmark for NLP anomaly detection with eight datasets and evaluations of nineteen state-of-the-art algorithms. Their findings highlight the superiority of transformer-based embeddings and reinforce the importance of data-driven methods in NLP anomaly detection. These data-driven approaches align with adaptive fault tolerance strategies in cloud environments, where real-time anomaly detection is critical for maintaining LLM stability.

\section{METHODOLOGIES}
\subsection{Failure prediction and dynamic resource allocation}
Fault prediction is at the heart of adaptive fault tolerance, which aims to intelligently warn and allocate resources in advance before failures occur. To achieve this, we use deep learning's multi-layer perceptron (MLP) model to predict failures by monitoring the system's performance metrics in real time. Set the performance index of the system state as $x_t=(x_1,x_2,\dots,x_n)$, where $x_i$ is the observed value of the $i$-th performance indicator at time $t$. Based on these observations, the neural network model predicts the system's failure probability $P(\text{fault}_t)$, which reflects the degree to which the current state of the system deviates from the normal range. The specific mathematical expression is Equation 1:

\begin{equation}
P(\text{fault}_t) = \sigma \left( \sum_{i=1}^{n} w_i \cdot x_{i,t} + b \right),
\label{eq:1}
\end{equation}

where $w_i$ is the weight of the $i$-th performance index, $b$ is the bias term, and $\sigma$ is the Sigmoid activation function. By training a neural network, the system is able to predict the probability of a failure based on current performance data. If $P(\text{fault}_t)$ exceeds the preset threshold $\theta$, the system will enter the fault warning state and start to adjust the resource allocation.

Based on the failure prediction results, the model dynamically adjusts the allocation of resources. Assuming that the current load of the system is $I_t$, the decision on the frequency of checkpoints $\lambda_t$ and the resource allocation strategy is given by Equation 2:

\begin{equation}
\lambda_t = \alpha \cdot P(\text{fault}_t) + \beta \cdot I_t,
\label{eq:2}
\end{equation}

where $\alpha$ and $\beta$ are the adjustment parameters, $I_t$ is the system load, and $P(\text{fault}_t)$ is the probability of the fault. This formula means that when the probability of system failure prediction increases or the load increases, the frequency of checkpoints also increases to ensure the stability of the system.

\subsection{Anomaly Detection and Fault Mitigation}
In order to deal with system anomalies in a timely manner, the proposed model combines anomaly detection algorithms to identify potential problems based on fault prediction. The change in the state of the system can be modeled by the Markov process, where the probability of the transfer of the state of the system between different time points is given by Equation 3:

\begin{equation}
P(s_{t+1} \mid s_t) = \frac{e^{-\lambda \cdot |s_{t+1} - s_t|}}{Z_t},
\label{eq:3}
\end{equation}

where $s_t$ and $s_{t+1}$ represent the system states at time $t$ and $t+1$, respectively, $\lambda$ is the attenuation factor, and $Z_t$ is the normalization constant. This formula describes the law of system state change; when the state changes greatly, the system is prone to failure or performance degradation, so it needs to pass through a fault-tolerant machine system to respond.

Once the system enters a high-risk state (i.e., $P(\text{fault}_t)$ exceeds the threshold), the system performs a failure mitigation action. In order to determine the most appropriate mitigation measures, an optimization objective function is introduced in this study for balancing between system load and failure impact. The objective function is Equation 4:

\begin{equation}
L(s_t) = \lambda_1 \cdot \text{ResourceCost}(s_t) + \lambda_2 \cdot \text{FaultImpact}(s_t),
\label{eq:4}
\end{equation}

where $\text{ResourceCost}(s_t)$ indicates the cost of current system resource consumption, $\text{FaultImpact}(s_t)$ is the impact of system state on performance, and $\lambda_1$ and $\lambda_2$ are tuning parameters. The purpose of the optimization goal is to select an appropriate resource allocation strategy to reduce the impact of system failures on performance while reducing resource overhead.

In the event of a failure, the system selects the most appropriate recovery measures based on the current state $s_t$ and the impact assessment of the failure. Assuming that the system can recover through state transition, the state transition probability is represented by the following Equation 5:

\begin{equation}
P(s_{t+1} \mid s_t, a_t) = \mathbb{E}[s_{t+1} \mid s_t, a_t],
\label{eq:5}
\end{equation}

where $a_t$ represents the operation performed at time $t$ (such as resource migration, checkpoint recovery, etc.), and $s_{t+1}$ is the system state at the next time.

By optimizing this formula, the system can quickly take appropriate action in the event of a failure, ensuring that resources are migrated and recovered in a timely manner, reducing system downtime.

To achieve efficient state migration and recovery, the model uses the following Equation 6 to select a standby resource $s_{\text{backup}}$ for failback:

\begin{equation}
s_{t+1} = s_{\text{backup}} \quad \text{if} \quad P(s_{t+1} \mid s_t, a_t) > \eta,
\label{eq:6}
\end{equation}

where $\eta$ is the set threshold, which indicates the stability standard that the system meets after migrating to a standby resource. When the system state is migrated to a standby resource, the model can ensure the speed and stability of fault recovery and minimize the impact on the operation of the model.

\section{EXPERIMENTS}
\subsection{Experimental setup}

he experiment utilizes a publicly-available dataset called the DSTC (Dialog State Tracking Challenges dataset), which is a widely used benchmark for the task of conversation state tracking and conversation management, especially in evaluating multi-turn conversational models. DSTC dataset consists of real-world task-oriented dialogues across various domains, such as restaurant and hotel booking or availability. The dataset is divided into multiple different dialogue scenarios, each scenario is divided into multiple rounds of conversation between users and the system, and the overall performance of the model in complex dialogue scenarios can be better evaluated. The dataset contains conversational data specifically selected to challenge the system's ability to track and understand conversational context, and highlights how well the system responds to evolving user goals throughout the conversation.

\subsection{Experimental analysis}
To fully evaluate the performance of the proposed adaptive fault tolerance mechanism, we chose four existing fault tolerance methods for comparison. Checkpointing (CP) \cite{kumari2021checkpointing} works by saving the state and intermediate results of a model periodically so that in case of a failure, computation can backtrack to the nearest checkpoint to recover, but performing frequent save operation may unnecessarily consume computational resources. Replica-based fault tolerance (RP) \cite{colomdistributed} increase redundancy be replicating the same tasks and data on multiple nodes and although the risk of a single point of failure reduces, it takes costly computing and storage resources. Another compared fault tolerance mechanism (SM) \cite{mudassar2022adaptive} based on state migration ensures the continuation of the task by transferring the execution status of the task to other available nodes, but its complexity of data synchronization and task scheduling is high. The anomaly detection with deep learning (AD) \cite{dhiman2021wind, li2024deception} method identifies potential anomalies by training deep learning models that are strongly adaptable and can be adjusted according to the real-time data of the system, but has a large dependence on the training data and models.

\begin{figure}[h!]
  \centering
  \begin{subfigure}[T]{1\linewidth}
    \includegraphics[width=\linewidth, height=0.7\linewidth]{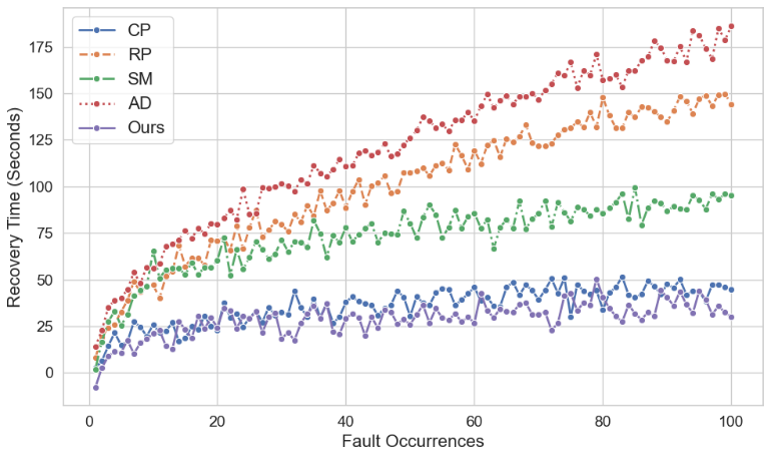}
  \end{subfigure}
  \caption{Comparison of Recovery Time for Different Methods}
  \label{fig:Comparison of Recovery Time for Different Methods}
\end{figure}

Comparison among various FTM approaches was performed based on recovery time which was taken as a key metric for our setup. Recovery time is the time taken by a system to recover to normal from a failed state after a failure has occurred. As shown in Figure \ref{fig:Comparison of Recovery Time for Different Methods}, Ours approach has much lower recovery time at all fault times compared with other approaches, demonstrating the effectiveness of our proposed adaptive fault tolerance mechanism in improving system reliability and reducing fault recovery time. In the case of high load or resource consumption, the Ours approach reduces system downtime and allows for speedy recovery by adapting recovery strategies on the fly. 

Figure \ref{fig:Comparison of Recovery Time for Different Methods} shows a comparison of different recovery practices with a fixed number of failures. where the abscissa represents the number of failures and the ordinate represents the time required to recover.

 We take fault prediction accuracy as the main evaluation index to evaluate the performance of various fault tolerance methods in fault prediction. The prediction accuracy of failure shows that how well each method can accommodate a range of fault scenarios, so the larger this measure; the closer the method to find out and predict failures for fault such as system.

\begin{figure}[h!]
  \centering
  \begin{subfigure}[T]{1\linewidth}
    \includegraphics[width=\linewidth, height=0.5\linewidth]{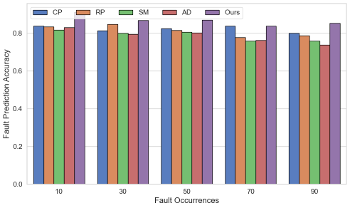}
  \end{subfigure}
  \caption{Fault Prediction Accuracy Comparison}
  \label{fig:Fault Prediction Accuracy Comparison}
\end{figure}

From results shown in Figure \ref{fig:Fault Prediction Accuracy Comparison}, Ours has better accuracy compared with others in all test cases, particularly in the high load and resources constraining cases, and have keeps steadily high accuracy at almost $90\%$ in failure predicting. By contrast, traditional CP, RP, SM, and AD methods tend to have low accuracy, accuracy decreases when the number of failures increases. The Ours method is the only one that was conscious of failures and responding to them according to its innovative adaptive fault tolerance mechanism for cloud computing environments, which indicates to what extent existing methods can predict failures but Ours is more important than existing methods. Additionally, computation cost is another essential indicator for fault tolerance. 

\begin{table}[h]
    \centering
    \renewcommand{\arraystretch}{1.5} 
    \setlength{\tabcolsep}{5pt}
    \begin{tabular}{|l|c|c|c|c|c|c|}
        \hline
        Methods & CP & RP & SM & AD & Ours \\
        \hline
        Computation Cost (Second) & 10.25 & 12.50 & 15.75 & 20.00 & 8.30 \\
        \hline
    \end{tabular}
    \caption{Computation Cost Comparison Results}
    \label{tab:Computation Cost Comparison Results}
\end{table}

Above Table \ref{tab:Computation Cost Comparison Results} shows the computation cost under 60 fault occurrences with 10 average calculations. From following Table \ref{tab:Computation Cost Comparison Results}, we can observe that our proposed method achieves lowest costs with existing methods.
\section{Conclusions}
In conclusion, we propose a novel adaptive fault tolerance mechanism to enhance the reliability and availability of large language models within cloud computing environments. By deriving the methods from the ground up, we show experimental results demonstrating that it is an improvement over the traditional fault-tolerant strategy for fault prediction accuracy and ultimately preserves the high reliability and high availability of the system at various loads. Notably, in the event of an increase in the number of faults, the Ours method maintains consistently high predictive accuracy, which shows significant advantages over currently available outliers detection methods. We can optimize our model even further in the future. Future studies could implement advanced machine learning methods like reinforcement learning for real-time fault-tolerant decision making to enhance the adaptability of the framework.

\renewcommand{\bibfont}{\footnotesize}

\footnotesize{
\bibliographystyle{IEEEtran}
\bibliography{main}
}

\end{document}